# Derivation of Relativistic law of Addition of Velocities from Superposition of Eigenfunctions and Discreteness

Mushfiq Ahmad

Department of Physics, Rajshahi University, Rajshahi, Bangladesh

E-mail: mushfiqahmad@ru.ac.bd

**Abstract**

We have defined a slowness, s, as the reciprocal conjugate of velocity, v. s = -ih/v. We have shown that Einstein's postulate (v has an upper limit) implies that s is discrete. A velocity operator is defined as the derivative with respect to s. Superposition of corresponding Eigenfunctions give Galilean law of addition of velocities. We have replaced the differential operator by the corresponding finite difference symmetric operator. We have shown that superposition of corresponding discrete Eigenfunctions give relativistic law of addition of velocities. A reciprocal symmetric number system is developed and with the help of this number system we have shown the relation between superposition and relativistic law of addition of velocities.

## 1. Introduction

We have already observed a reciprocal symmetry and correspondence between quantum mechanics and special relativity[1]. In this paper we intend to study in detail the relation between the Principle of Superposition of Einstein's law of addition of velocities. This we shall study in Part I. In Part II we shall study the how Reciprocal Symmetry relates superposition of Eigenfunctions to relativistic addition of velocities. In Part III we shall develop reciprocal symmetric number system and study how it is related to the results of the preceding parts.

## PART I
### Velocity Eigenfunction and Relativistic Addition of Velocities

We define a "slowness", s, as the reciprocal conjugate of velocity, v, by the relation sv = -ih. Like a momentum operator[2] we then define a velocity operator and find the corresponding Eigenfunction. Superposition of two Eigenfunctions corresponding to velocities u and v gives the Eigenfunction corresponding to u+v. This is Galilean addition of velocities.

Einstein's postulate sets an upper limit, c, to v. This implies a lower limit to s. We call it q = -ih/c. We replace the differential velocity operator by the corresponding to finite difference operator and we set q as the lower limit of s. Superposition of velocities Eigenfunctions should give Einstein's law of addition of velocities. We shall study this in section 2.

## 2. Reciprocal Velocity and Velocity Operator

We define slowness, s, as the reciprocal conjugate of v

$$s = -i\hbar/v \tag{2.1}$$

And q is the reciprocal conjugate of, c, the speed of light

$$q = -i\hbar/c \tag{2.2}$$

The velocity operator is

$$-i\hbar\frac{\partial}{\partial s} \tag{2.3}$$

The Eigenvalue equation is

$$-i\hbar\frac{\partial}{\partial s}\psi = v\psi \tag{2.4}$$

## 3. Superposition of Continuous Eigenfunctions and Galilean Addition of Velocities

Eigenfunction $\psi(v) = \exp(is.v/\hbar)$ gives the Eigenvalue v

Superposition of Eigenfunctions gives the resultant Eigenfunction

$$\psi(u)\psi(v) = \psi(u+v) \tag{3.1}$$

$u + v$ is the Galilean sum of velocities $u$ and $v$.

## 4. Discrete Eigenfunctions and Lorentz-Einstein Addition of Velocities

We now replace the continuous Eigenvalue equation by the corresponding to discrete equation

$$-i\hbar \frac{D}{\partial D(s,q)} \phi = v\phi \qquad (4.1)$$

Where

$$\frac{D}{\partial D(s,q)} \phi = V(v) \frac{\phi(s+q) - \phi(s-q)}{2q} \qquad (4.2)$$

We require

$$\left. \begin{array}{c} \dfrac{D}{D(s,q))} \to \dfrac{\partial}{\partial s} \\ V(v) \to v \\ \phi \to \psi \end{array} \right\} \text{as } q \to 0 \qquad (4.3)$$

We also require that (4.2) remains invariant under the change $q \to -q$. The relation between negation and reciprocation is studied in Appendix ….

This requirement implies the requirement

$$\left. \begin{array}{c} V(v) \to V(v) \\ \phi \to \phi \end{array} \right\} \text{as } q \to -q \qquad (4.4)$$

Our requirements are fulfilled if

$$V(v) = \frac{v}{\sqrt{1 + (qv/\hbar)^2}} = \frac{v}{\sqrt{1 - (v/c)^2}} \qquad (4.5)$$

And

$$\phi(v) = \left( \frac{1 + iqv/\hbar}{\sqrt{1 + (qv/\hbar)^2}} \right)^{s/q} = \left( \frac{1 + v/c}{\sqrt{1 - (v/c)^2}} \right)^{s/q} \qquad (4.6)$$

Relation (4.2) has been used to get the r.h.s of the above relations.

Superposition of 2 Eigenfunctions gives

$$\phi(u).\phi(v) = \phi(w) = \left(\frac{1+iqw/\hbar}{\sqrt{1+(qw/\hbar)^2}}\right)^{s/q} \tag{4.7}$$

Where

$$w = \frac{u+v}{1-u.v(q/\hbar)^2} = \frac{u+v}{1+uv/c^2} \tag{4.8}$$

This is Lorentz-Einstein sum of velocities[3]

# PART II

## Reciprocal Symmetry in Quantum Mechanics and Special Relativity

### 5. Generalized Reciprocal and Negative

*Reciprocal*

$$a \oplus_0 b = \frac{a+b}{1+ab} \tag{5.1}$$

[Compare (5.1) to (15.3) when $\varphi = 0$]. We observe that the "sum" (5.1) is reciprocal symmetric i.e.

$$(1/a) \oplus_0 (1/b) = a \oplus_0 b \tag{5.2}$$

We also observe that the product

$$A \oplus_1 B = A.B \tag{5.3}$$

is symmetry under negation i.e.

$$(-A) \oplus_1 (-b) = A \oplus_1 B \tag{5.4}$$

Introducing the reciprocal operator $R$, which takes $a$ to its reciprocal $R(\varphi, a)$ we write [see (15.2) for $c=1$]

$$R(\varphi = 0, a) = 1/a \text{ and } R(\varphi = 1, a) = -a \tag{5.5}$$

We can write (5.2) and (5.4) under the same general relation

$$R(\varphi, a) \oplus_\varphi R(\varphi, b) = a \oplus_\varphi b \tag{5.6}$$

$\varphi = 0$ in (5.6) will give (5.2), while $\varphi = 1$ in (5.6) will give (5.4).

[for mathematical details see B11 of section 19]

# 6. Reciprocal Symmetry

*Reciprocal Symmetry*

We observe that (4.7) and (4.8) remain invariant under general reciprocal transformations

$$\{-\phi(u)\}.\{-\phi(v)\} \qquad (6.1)$$

and

$$w = \frac{u+v}{1+uv/c^2} = \frac{(c^2/u)+(c^2/v)}{1+\{(c^2/u).(c^2/v)\}/c^2} \qquad (6.2)$$

Choosing $s/q=2$ in (4.6) we get

$$\phi(v) = \left(\frac{1+v/c}{\sqrt{1-(v/c)^2}}\right)^2 = \frac{1+v/c}{1-v/c} \qquad (6.3)$$

Therefore

$$\varphi(c^2/v) = -\phi(v) \qquad (6.4)$$

Using definitions (5.7) and also introducing the notation $\varphi(v) = \varphi'(v/c)$

$$\phi'(R(\varphi=0, v/c)) = R(\varphi=1, \phi') \qquad (6.5)$$

Therefore Einstein's law of addition of velocities and velocities Eigenfunctions demonstrate the same reciprocal symmetry.

*"Shift" of Origin*

Transformation of number $a_0$, measured in the system in which the neutral number is 0, to $a_0$, in the system in which the neutral number is 1, is given by transformation relation.

$$a_\varphi = \frac{a_0 + \varphi}{1 - a_0.\varphi} \qquad (6.6)$$

[for details see (14.2) below].

for $s/q = 2$ and $a_0 = v/c$ we have

$$\phi = a_1 = \frac{1+v/c}{1-v/c} \qquad (6.7)$$

## PART IV

## Reciprocal Symmetric Real Number System

### 7. Axioms of Real Number System

Consider the Field Axioms[4] satisfied by the members of the set, **R**, of real numbers.

A1. $a+b=b+a$

A2. $(a+b)+d=a+(b+d)$

A3. There exists a 0 in **R** such that $\quad a+0=a$ for all $a$ in **R**

A4. For every $a$ there is a $N(0,a)$ in **R** such that $\quad a+N(0,a)=0$

A'1. $a.b=b.a$

A'2. $(a.b).d=a.(b.d)$

A'3. There exists a 1 in **R** such that $\quad a.1=a$ for all $a$ in **R**

A'4. For every a there is a $N(1,a)$ in **R** such that $\quad a+N(1,a)=0$

We observe that the two sets of axioms are essentially the same except the difference in 0 and 1 in A3 and A'3.

*Axioms of Extended Real Numbers*

We extend the real number system by including 2 elements $+\infty$ and $-\infty$ to the set of real numbers. This set, **E**, will be called the set of extended real numbers[5]. The following set of axioms is included.

E1. $a+\infty=\infty$

E2. $a-\infty=-\infty$

E3. $a.\infty=\infty$ if $a>0$

E4. $a.-\infty=-\infty$ if $a\neq\infty$

For all real number a

E4. $\infty - \infty$ is not defined

E5. $0.\infty$ is not defined

## 8. Generalized Notation

Introducing the notation $+ = \oplus_0$ and $. = \oplus_1$ the above sets of axioms may be rewritten as

A1. $a \oplus_0 b = b \oplus_0 a$

Etc. and

A'1. $a \oplus_1 b = b \oplus_1 a$

Introducing the unspecified number $\varphi$ which, may be allowed to take values $\varphi = 0$ and $\varphi = 1$, the above sets of axioms may be written as

A"1. $a \oplus_\varphi b = b \oplus_\varphi a$

A"2. $(a \oplus_\varphi b) \oplus_\varphi = a \oplus_\varphi (b \oplus_\varphi d)$

A"3. There exists a $a \oplus_\varphi \varphi = a$ in $R$ such that $a \oplus_\varphi \varphi = a$

A"4. There exists a $N(\varphi, a)$ for every a in $R$ such that $a \oplus_\varphi N(\varphi, a) = \varphi$

The set A" of axioms is just the sets A and A' rewritten. To achieve a real generalization, we re-enunciate the axioms as below.

## 9. Generalization of Addition to any Neutral Number $\varphi$

B1. It is possible to define $a \oplus_\varphi b$ (depending on an arbitrarily chosen $\varphi$ in $R$) for any $a$ and $b$ in $R$ such that $a \oplus_\varphi b$ belongs to $R$ and that $a \oplus_\varphi b = b \oplus_\varphi a$

B2. $(a \oplus_\varphi b) \oplus_\varphi = a \oplus_\varphi (b \oplus_\varphi d)$

B3. There exists a $a \oplus_\varphi \varphi = a$ in $R$ such that $a \oplus_\varphi \varphi = a$

B4. There exists a $N(\varphi, a)$ for every a in $R$ such that $a \oplus_\varphi N(\varphi, a) = \varphi$

## 10. Reflection Axioms

B5. Corresponding to every $a$ in $\boldsymbol{R}$ there exists, depending upon an arbitrarily chosen $c \neq \varphi$, a $R(\varphi, a)$ in $\boldsymbol{R}$.

B6. $N(R(\varphi, a)) = R(N(\varphi, a))$

B7. $R^2(\varphi, a) = a$ and $N^2(\varphi, a) = a$

B8. $N(\varphi, \varphi) = \varphi$

B9. $R(\varphi, c) = c$

B10. $N(\varphi, a) \oplus_\varphi N(\varphi, b) = N(\varphi, a \oplus_\varphi b)$

Theorem 1: $a \oplus_\varphi N(\varphi, a) = \varphi$

Theorem 1 above is Axiom B4. Therefore, it is not necessary to put B4 as a separate axiom. We keep it, nevertheless, for comparison with axioms A and A'.

## 11. Symmetry Axiom

B11. $R(\varphi, a) \oplus_\varphi R(\varphi, b) = a \oplus_\varphi b$

Note: B11 contains the equivalent of E1 (see Theorem 3).

Theorem 2: $a \oplus_\varphi R(\varphi, b) = R(\varphi, a \oplus_\varphi b)$

(See Appendix 2)

Theorem 3: $a \oplus_\varphi c = c$

(See Appendix 3)

## 12. Multiplication Axioms

Greek letters $\alpha, \beta, \gamma$ etc. stand for elements of the familiar set, $\boldsymbol{F}$, of real numbers extending from $-\infty$ to $+\infty$. Since there is no risk of confusion, we shall write $\otimes$ for $\otimes_\varphi$ and $\oplus$ for $\oplus_\varphi$ (omitting the subscript $\varphi$).

B12. $\alpha \otimes a$ belongs to **R** for every $\alpha$ in **F** and $a$ in **R**

B13. $\alpha \otimes (a \oplus b) = (\alpha \otimes a) \oplus (\alpha \otimes b)$

B14. $(\alpha \otimes a) \oplus (\beta \otimes a) = (\alpha + \beta) \otimes a$ where + is the familiar addition

B15. $\alpha \otimes (\beta \otimes a) = (\alpha.\beta) \otimes a$ where . is the familiar multiplication.

B16. $(-\alpha) \otimes a) = \alpha \otimes N(a) = N(\alpha \otimes a)$ where -a is the familiar negative of a.

B17. $0 \otimes a = \alpha \otimes \varphi = \varphi$.

Axioms B12-B17 also ensure consistency between Homogenous number system and the familiar number system.

*Undefined Quantities*

U1. $c \oplus N(c)$ is not defined

U2. $0 \otimes c$ is not defined

U3. $\infty \otimes \varphi$ is not defined

Note: U1 corresponds to E4. U2 and U3 correspond to E5.

### 13. Measures and Counts

The numbers we are talking about in sections 18, 19 etc., we shall call measures. On the other hand, the numbers for which we are employing Greek alphabets in section 20, we shall call counts.

### 14. Isomorphic Transformations

Let $a_\varphi$ and $a_\theta$ be the numbers which represent the same physical quantity; the first in the system in which $\varphi$ is the neutral number, the second in the system in which $\theta$ is the neutral number (the origin). $a_0$ will, then, be the number in the familiar number system measured from 0. The transformation relation relating two numbers is

$$a_\varphi = \frac{(1+\theta.\varphi).a_\theta - (\theta - \varphi)}{(1+\theta.\varphi) + a_\theta.(\theta - \varphi)} \qquad (14.1)$$

Transformation to $a_\varphi$ from the familiar number $a_0$ will be

$$a_\varphi = \frac{a_0 + \varphi}{1 - a_0 . \varphi} \qquad (14.2)$$

## 15. Representations

$$N(\varphi, a) = \frac{(\varphi - a) + (1 + a.\varphi).\varphi}{(1 + a.\varphi) - (\varphi - a)\varphi} \qquad (15.1)$$

$$R(\varphi, a) = \frac{(c - \varphi)^2 (1 + a.\varphi) + \varphi(1 + c.\varphi)^2 (a - \varphi)}{(1 + c\varphi)^2 (a - \varphi) - (c - \varphi)^2 (1 + a.\varphi)\varphi} \qquad (15.2)$$

$$a \oplus b = \frac{y + \varphi}{1 - y\varphi} \quad \text{where} \quad y = \frac{\dfrac{a - \varphi}{1 + a\varphi} + \dfrac{b - \varphi}{1 + b\varphi}}{1 + \left(\dfrac{1 + c\varphi}{c - \varphi}\right)^2 \dfrac{a - \varphi}{1 + a\varphi} \dfrac{b - \varphi}{1 + b\varphi}} \qquad (15.3)$$

$$\alpha \otimes a = \frac{\left(\dfrac{c - \varphi}{1 + c\varphi}\right) \dfrac{w^\alpha - 1}{w^\alpha + 1} + \varphi}{1 - \left(\dfrac{c - \varphi}{1 + c\varphi}\right) \dfrac{w^\alpha - 1}{w^\alpha + 1} \varphi} \quad \text{where} \quad w = \frac{\dfrac{c - \varphi}{1 + c\varphi} + \dfrac{a - \varphi}{1 + a\varphi}}{\dfrac{c - \varphi}{1 + c\varphi} - \dfrac{a - \varphi}{1 + a\varphi}} \qquad (15.4)$$

*Special Cases*

When $\varphi = 0$ and $\varphi = 1$ respectively

The general negatives are:

$$N(\varphi = 0, a) = -a \quad \text{and} \quad N(\varphi = 1, a) = 1/a \qquad (15.5)$$

The general reciprocals are: When $c = 1$ it becomes

$$R(\varphi = 0, a) = \frac{1}{a} \quad \text{and} \quad R(\varphi = 1, a) = -a \qquad (15.6)$$

$$R(0, a) = \frac{1}{a} \qquad (15.7)$$

In this case the sum $\oplus_\varphi$ becomes $\oplus_0$ and

$$a \oplus_0 b = \frac{a+b}{1+\dfrac{a.b}{c^2}} \qquad (15.8)$$

(15.8) has the form of Lorentz-Einstein law of addition of velocities.

With $\varphi = 0$ and $c \to \infty$ the multiplication becomes

$$a \otimes_0 a \xrightarrow[c \to \infty]{} a.a \qquad (15.9)$$

With $\varphi = 1$ and $c = 0$ In this case negation becomes

$$N(\varphi, a) = N(1, a) = \frac{1}{a} \qquad (15.10)$$

The reciprocal is

$$R(\varphi, a) = R(1, a) = -a \qquad (15.11)$$

the sum becomes

$$a \oplus_1 b = a.b \qquad (15.12)$$

## 16. Addition with General Negative

*Theorem:*

$$a \oplus_\varphi N(\varphi, a) = \varphi \qquad (16.0a)$$

When there is no risk of ambiguity, we shall put
$N(\varphi, a) = N(a)$ and $R(\varphi, a) = R(a)$.

With this notation we rewrite the theorem

*Theorem:*

$$a \oplus N(a) = \varphi \qquad (16.0)$$

*Proof:* Let

$$a \oplus N(a) = y \qquad (16.1)$$

Using B10

$$N(a) \oplus N^2(a) = N(y) \qquad (16.2)$$

Using B7 and B1

$$a \oplus N(a) = N(y) \tag{16.3}$$

Comparing (16.1) and A(1,3)

$$y = N(y) \tag{16.4}$$

Therefore

$$y = \varphi \text{ or } R(\varphi) \tag{16.5}$$

Two definitions of $\oplus$ are possible. One will give $a \oplus N(a) = \varphi$. The other one will give $a \oplus N(a) = R(\varphi)$. We shall use the notation $\oplus_\varphi$ when it gives $\varphi$. We shall use the notation $\oplus_{R(\varphi)}$ when the neutral number is $R(\varphi)$.

## 17. Addition with General Reciprocal

*Theorem:*

$$a \oplus R(b) = R(a \oplus b) \tag{17.0}$$

*Proof:* Consider the sum

$$N(b) \oplus d = a \tag{17.1}$$

By B11 and B7

$$N(b) \oplus d = a = R(N(b)) \oplus R(d) = a \tag{17.2}$$

Therefore, using B4 and $N(R(N(b))) = R(b)$ right hand part of (17.2) gives

$$a \oplus R(b) = R(d) \tag{17.3}$$

Again using left hand part of (17.2) and $N^2(b) = b$

$$a \oplus b = d \tag{17.4}$$

Comparing (17.3) and (17.4)

$$a \oplus R(b) = R(a \oplus b) \tag{17.5}$$

## 18. Reciprocal Symmetric Analogue of Einstein's Postulate

*Theorem:*

$$a \oplus c = c \qquad (18.0)$$

*Proof:* Let

$$a \oplus c = d \qquad (18.1)$$

Using B11 and B9

$$R(a) \oplus c = d \qquad (18.2)$$

Also using (17.0) and $R(c) = c$

$$R(a) \oplus c = R(d) \qquad (18.3)$$

Therefore,

$$R(d) = d \qquad (18.4)$$

Therefore,

$$d = c \qquad (18.5)$$

Therefore,

$$a \oplus c = c \qquad (18.6)$$

## 28. Conclusion

In Part I we have defined reciprocal conjugate variables, and we have been able to derive relativistic law of addition of velocities from superposition of discrete wave functions.

In Part II we have shown that both quantum mechanics principle of superposition and relativistic law of addition of velocities satisfy the same generalized reciprocal symmetry

In Part IV we have given algebraic details of reciprocal symmetric real number system.